# Electric-field controlled superconductor-ferromagnetic insulator transition

L. K. Ma[1*], B. Lei[1*], N. Z. Wang[1*], K. S. Yang[2], D. Y. Liu[2], F. B. Meng[1], C. Shang[1], Z. L. Sun[1], J. H. Cui[1], C. S. Zhu[1], T. Wu[1,5,6], Z. Sun[3,6], L. J. Zou[2] and X. H. Chen[1,4,5,6]

1. Hefei National Laboratory for Physical Sciences at Microscale and Department of Physics, and CAS Key Laboratory of Strongly-coupled Quantum Matter Physics, University of Science and Technology of China, Hefei, Anhui 230026, China

2. Key Laboratory of Materials Physics, Institute of Solid State Physics, Chinese Academy of Sciences, Hefei 230031, Anhui, China

3. National Synchrotron Radiation Laboratory, University of Science and Technology of China, Hefei, Anhui 230026, China

4. CAS Center for Excellence in Superconducting Electronics (CENSE), Shanghai 200050, China

5. CAS Center for Excellence in Quantum Information and Quantum Physics, Hefei, Anhui 230026, China

6. Collaborative Innovation Center of Advanced Microstructures, Nanjing University, Nanjing 210093, China

* These authors contributed equally to this work

**Abstract:**

How to control collectively ordered electronic states is a core interest of condensed matter physics. We report an electric field controlled reversible transition from superconductor to ferromagnetic insulator in (Li,Fe)OHFeSe thin flake using solid ion conductor as the gate dielectric. By driving Li ions into and out of the (Li,Fe)OHFeSe thin flake with electric field, we obtained a dome-shaped superconducting region with optimal $T_c$ ~ 43 K, which is separated by a quantum critical point from ferromagnetically insulating phase. The ferromagnetism arises from the long range order of the interstitial Fe ions expelled from the (Li,Fe)OH layers by Li injection. The device can reversibly manipulate collectively ordered electronic states and stabilize new metastable structures by electric field.

**Main text:**

The manipulation of collectively ordered electronic states is always the issue of long-standing focus in condensed matter physics, many exotic electronic phases have been discovered with exertion of various methods including chemical doping and applying high pressure [1-4]. As the most widely utilized approach, chemical doping has triggered the finding of high-$T_c$ superconductors by suppressing antiferromagnetism or spin density wave via introducing charge carriers in cuprates and iron-based superconductors, respectively [1-3]. Moreover, chemical doping could also induce new structures, and consequently lead to novel electronic states. Nevertheless, the doping range is usually limited by the chemical solubility of dopants and leaves many exotic phases unexplored. Another approach to precisely control charge carrier concentration is the gating technique based on field-effect transistors (FETs) [5-7]. Through the electric-field-induced electrostatic doping with dielectric or electric-double-layer (EDL) surface gating via ionic liquid, electronic states of a two-dimensional system can be tuned continuously, and new electronic states could be realized beyond the limitation of conventional chemical doping [8-10]. However, these techniques could only tune the carrier density for thin flakes and the tuning depth is restricted to a few nanometers beneath the surface due to the Thomas-Fermi screening [11]. Recently, using solid ion conductor (SIC) as the gate dielectric, we have developed a new type of FET device, namely SIC-FET [12-14]. Using the lithium ion conductive glass ceramics as the gate dielectric, Li ions can be driven into FeSe thin flakes, and enhance the $T_c$=8 K in pristine FeSe to a maximum of 46 K. In addition, a new structural transition induces $Li_xFe_2Se_2$ crystalline phase out of the pristine FeSe phase, which is stabilized by electric field and not accessible by conventional methods [12].

Here we demonstrate that we successfully tune the electronic properties of (Li,Fe)OHFeSe thin flakes with thickness of 120 nm by electric field using the SIC-FET device. Electric field control of magnetism is attracting the accumulative interest and could be potentially utilized for practical applications [15-20]. High-$T_c$ superconductivity usually develops from antiferromagnetic insulators, and the high-$T_c$ superconductivity and ferromagnetism are mutually exclusive [21,22]. It is striking that a reversible transition from high-$T_c$ superconductor to ferromagnetic insulator can be realized by driving Li ions into and out of the (Li,Fe)OHFeSe thin flake with electric field. A dome-shaped superconducting phase diagram is observed with Li doping, and $T_c$ increases from 27 K to 43 K at the optimal doping. In the phase diagram, there exists a quantum critical point, which separates high-$T_c$ superconductor and ferromagnetic insulator. The latter arises from that Li ions driven into the thin flake by electric field replace Fe ions in the (Li,Fe)OH layers, and the Fe ions migrate to the interstitial sites and becomes ordered. Which leads to a long-range ordering ferromagnetism and a structural transformation to $Fe_xLiOHFeSe$ crystalline phase. When the gate voltage is switched back, the ferromagnetic insulating phase fully reverses to the initial superconducting phase, indicating that we can manipulate the collectively ordered electronic state by the electric field.

Figure 1A is a schematic illustration of the SIC-FET device used in our studies. The detailed device fabrication is described in the supplementary materials. The exfoliated (Li,Fe)OHFeSe thin flakes with typical thickness of ~120 nm are used to fabricate the transport channel. The inset of Fig. 1B shows the optical image of a (Li,Fe)OHFeSe thin flake with a standard Hall bar configuration and with current and voltage terminals labeled. Li ions in the lithium ion conductor can be precisely controlled and driven into the thin flakes by electric field. Fig. 1B shows a typical $R$-$V_g$ curve with a continuously sweeping rate of 1 mVs$^{-1}$ at $T$ = 260 K. The resistance of the sample remains almost unchanged with gating voltage for $V_g$ < 4.25 V, and starts to drop evidently at $V_g$ = 4.25 V, and reaches a minimum around $V_g$ = 4.9 V, then increases drastically. When the gate voltage is swept back, the resistance continuously increases rapidly and then falls down. At $V_g$ = -2 V, the resistance returns to a value close to the initial state. This behavior indicates that the gate tuning process is highly reversible.

Figures 1C and 1D show the temperature dependent resistance of (Li,Fe)OHFeSe thin flake at various gate voltages. At $V_g$ = 0 V, the (Li,Fe)OHFeSe thin flake shows superconductivity with an onset transition temperature $T_c$ = 27 K. With increasing the gate voltage, Li ions are gradually driven into the sample, accompanied by an increase of $T_c$. When $V_g$ = 4.91 V, the optimal superconductivity is achieved with $T_c$ = 43 K, which is the same as the highest $T_c$ in polycrystalline $(Li_{0.8}Fe_{0.2})OHFeSe$

[23]. With further increasing the Li content, $T_c$ gradually decreases, and eventually the sample becomes an insulator. When the gate voltage is swept back to −2 V, both of the resistance and $T_c$ recover and are close to the initial state before gating as shown in Fig. 1C. The difference between them indicates that the Li injection-extraction cycle slightly changes microstructures, however, the cycle controlled by electric field is highly repeatable in the measurements.

To further reveal the evolution of electronic properties in the gating process, the magnetic-field ($H$) dependent Hall resistance $R_{xy}$ at $T = 60$ K is measured at different gating voltages. Figs. 2A and 2B show a linear magnetic-field dependence of *Hall* resistance in the superconducting regime. In contrast, it shows a superposition of a linear and a square-shaped hysteretic dependence in the insulating state as shown in Fig. 2C, indicating an anomalous *Hall* effect in the insulating state. Figs. 2D and 2E show the magnetic field dependence of *Hall* resistance and magnetoresistance at different temperatures. In Fig. 2D, the square-shaped hysteretic *Hall* resistance indicates that the sample is tuned to be a *ferromagnetic* insulator below 175 K. In the ferromagnetic state, the total *Hall* resistance $R_{xy}$ can be expressed as $R_{xy} = R_A M + R_H H$ [24,25], where $R_A$ and $R_H$ are the anomalous and ordinary *Hall* coefficients, respectively. $M$ is the magnetization of the sample, and $H$ is the external magnetic field. Fig. 2E illustrates the magnetic field dependent magnetoresistance (*MR*) measured at the corresponding temperatures, and a clear butterfly-shaped hysteresis arises, which is due to the spin-dependent scattering of carriers by local magnetic ordering. These results indicate the existence of a long-range ferromagnetic order in the insulating phase. The peak position in *MR* corresponds to the coercive field ($H_c$), and the reduced scattering of a specific spin orientation leads to negative *MR* on either side of $H_c$ [25]. In both $R_{xy}$-$H$ and *MR* curves, the smaller $H_c$ at higher temperature suggests that the ferromagnetic order becomes more weak with increasing temperature, which is a common characteristics in ferromagnetic materials [24]. It should be pointed out that these measurements are limited to the temperature below $T = 175$ K because the lithium ions could not be confined for $T > 175$ K. Moreover, our data suggest that the transition temperature of ferromagnetism (Curie temperature) should be much higher than 175 K, which cannot be determined in our measurements. In addition, the angular dependence of the *MR* is measured (see fig. S4 in the supplementary materials). We find that the coercive field ($H_c$) increases with increasing the angle ($\theta$) between the direction of the applied magnetic field and the c-axis. As shown in fig. S4 in supplementary materials, the coercive field is proportional to $1/\cos\theta$ (that is, $H_c \sim 1/\cos\theta$). This behavior demonstrates that the easy axis of magnetization is along the c axis of (Li,Fe)OHFeSe with very strong anisotropy. When the gate voltage is swept back to −2 V, the magnetic field dependent *Hall* resistance recovers to a linear behavior, similar to the $R_{xy}(H)$ before gating without any anomaly (see fig. S5B in the supplementary materials). All these results indicate a superconducting-ferromagnetic insulating transition in (Li,Fe)OHFeSe with SIC-FET device, which can be reversibly controlled by electric field.

*In-situ* X-ray diffraction (XRD) is performed on the SIC-FET device at $T = 150$ K to study the structural evolution of (Li,Fe)OHFeSe thin flake with gating as shown in Fig. 3. We find that the injection of Li ions to the thin flake by electric field leads to structural modifications. The (001) diffraction peak of (Li,Fe)OHFeSe locates at $2\theta = 9.56°$ and $9.66°$ at $T = 300$ K and 150 K before gating, respectively. With Li ions being driven into the thin flake by electric field, the (001) peak shows no noticeable variation in intensity and position until the gate voltage increases up to 4.91 V, at which the optimal $T_c$ of 43 K is achieved. With further increasing the gate voltage, a new diffraction peak appears at lower angle of 8.44° (corresponding *d*-value of 10.48 Å), and it becomes stronger and moves towards lower angle with increasing Li ions driven into the thin flake. Concomitantly, the intensity of original (001) peak of (Li,Fe)OHFeSe gradually decreases. This result indicates a structural transformation from (Li,Fe)OHFeSe phase to a new structural phase induced by Li injection. When the gating voltage is increased to 5.24 V, the (Li,Fe)OHFeSe phase completely disappears as shown in Fig. 3B. We note here that the coexistence of the (Li,Fe)OHFeSe phase and the new structural phase could be attributed to the inhomogeneity of the Li ions distribution. Such inhomogeneous distribution of Li ions in the sample is evidenced by the fact that the lattice parameter of c-axis for (Li,Fe)OHFeSe phase remains the same when the new structural phase shows up and grows as shown in Fig. 3. In addition, the inhomogeneous distribution of Li ions in the sample strongly depends on the thickness of the thin flake. It is possible that there is no coexistence of the two structural phases in a thinner flake. We have tried to perform XRD measurements on the device with the flake thickness less than 150 nm, but the sample is too thin (and too small) to yield decent diffraction signal.

We stress that there exist two kinds of evolution in X-ray diffraction patterns during gating as the gate voltage is sweeping backwards to −2 V as shown in Fig. 3 and in fig. S2, respectively. The former occurs in an irreversible gating process, while the latter takes place in a reversible process. In Fig. 3, the (001) diffraction peak of the new structural phase shifts from 8.02° to 8.92° (corresponding *d*-value from 11.02 Å to 9.91 Å) when the gate voltage varies from 5.29 V to −2 V. The *d*-value of 9.91 Å is much larger than that (9.21 Å) of the original (Li,Fe)OHFeSe phase. It suggests that the gating process is irreversible. Such an irreversibility is also confirmed in electric transport. As shown in Fig. 3C, the resistance is one order magnitude larger than that before gating, and it shows insulating behavior after the gate voltage is swept to −2 V as shown in fig. S5C. However, in the reversible gating process, the diffraction peak of the new structural phase shifts from 8.21° to 9.35° (corresponding $d$-value from 10.77 Å to 9.46 Å) when the gate voltage is swept from 5.14 V to −2 V as shown in fig. S2, and the $d$-value of 9.46 Å is close to that (9.21 Å) of the original (001) diffraction peak of (Li,Fe)OHFeSe phase. This behavior indicates that the gating process is nearly reversible, which is evidenced both by the gating curve in fig. S2C and by the recovery of superconductivity with nearly the same $T_c$ from a ferromagnetic insulating state in fig. S5A. These results indicate that the newly formed structural phase is stabilized only by electric field, and that whether the gating process controlled by electric field is reversible or not strongly depends on the maximal voltage applied in the gating process. As will be shown later, we propose a scenario of the lithiation process to explain these striking properties.

Based on the transport and *in-situ* X-ray diffraction measurements, the phase diagram for the gate-voltage tuned (Li,Fe)OHFeSe thin flake is plotted in Fig. 4. At $V_g = 0$ V, the (Li,Fe)OHFeSe thin flake shows superconductivity with $T_c$ =27 K. As listed in Table S1 of the supplementary materials, the composition of the thin flake obtained by structural refinement is (Li$_{0.81}$Fe$_{0.19}$)OHFe$_{0.97}$Se. The existence of Fe vacancies in the selenide layers is responsible for the low-$T_c$ of 27 K [26,27]. With increasing the gate voltage, Li ions are gradually driven into the thin flake, and $T_c$ shows a dome-like behavior. The optimal superconductivity with $T_c$ = 43 K is achieved when the gate voltage is increased to 4.91 V, which is the same as that of (Li$_{0.8}$Fe$_{0.2}$)OHFeSe [23].

With further increasing the gate voltage, $T_c$ gradually decreases, and then the superconductivity is completely suppressed, and eventually the system goes into a ferromagnetic insulator. It is striking that there exists a quantum critical point at the gating voltage of 5.13 V, which separates superconducting and ferromagnetic insulating state. As shown in Fig. 4, the lattice parameter of $c$-axis for (Li,Fe)OHFeSe phase monotonically increases with increasing the gate voltage up to 4.91 V, then saturates with further increasing the gate voltage, accompanied by the appearance of a novel structural phase with (001) diffraction peak at $2\theta=8.44°$. The $c$-axis lattice parameter of the novel structural phase monotonously increases with increasing the gate voltage. There exists a boundary between reversibility and irreversibility for the gating process around $V_g = 5.18$ V, and the gating process is irreversible for $V_g > 5.18$ V. It should be addressed that coexistence of high-$T_c$ superconductivity and ferromagnetic state in the same phase diagram is observed for the first time.

We propose a scenario of the lithiation process controlled by electric field to explain these findings in (Li,Fe)OHFeSe thin flake, which is also supported by first-principles calculations as will be shown later. Fe in the (Li,Fe)OHFeSe phase has two different crystallographic positions, Fe1 is in the selenide layers and Fe2 is in (Li,Fe)OH layers. The Li ions driven into (Li,Fe)OHFeSe thin flake prefer to enter into the (Li,Fe)OH layer and replace the Fe2. As listed in Table S1 in the supplementary materials, the starting material is $(Li_{0.81}Fe_{0.19})OHFe_{0.97}Se$ with $T_c = 27$ K. When the Li ions are initially driven into the thin flake up to $x = 0.03$, Li ions replace the Fe in the hydroxide layers and the Fe ions expelled by Li can migrate away from the hydroxide layers to fill the vacancies in the selenide layers. Once the vacancies are filled, the thin flake achieves the optimal $T_c \sim 43$ K. Similar lithiation process has been observed in hydrothermally synthesized (Li$_{1-}$

$_x$Fe$_x$)OHFe$_{1-y}$Se polycrystalline powder lithiated with *n*-BuLi lithiating reagent [27]. The composition of the thin flake can be considered as Fe$_{x-0.03}$(Li$_{0.81+x}$Fe$_{0.19-x}$)OHFeSe when Li ions are further driven into the thin flake by electric field. With more Li ions ($x > 0.03$), the Fe ions extruded from the hydroxide layers migrate to the interstitial sites, which is evidenced by the appearance of the new structural diffraction reflection just after the optimal doping is achieved as shown in Fig. 3. When the amount of the aggregated Fe reaches a certain value, the interstitial Fe ions become ordered, and eventually lead to a long-range ferromagnetic order. When $V_g$ is swept back to -2 V, Li ions are driven out of the hydroxide layers, and Fe ions at the interstitial sites can re-occupy the original positions in (Li,Fe)OH layer, leading to the disappearance of ferromagnetic state and the recovery of superconductivity as shown in figs. S5A and S5B of the supplementary materials. When the Fe in the hydroxide layer is completely replaced by Li, the composition of the thin flake is Fe$_{0.16}$LiOHFeSe, which is a ferromagnetic insulator. When Li ions are further driven into the thin flake, there exist two possible cases. One is that Li may expel the Fe from the selenide layer as pointed out by Woodruff et al. [27]. Another is that further lithiation cannot extrude the Fe from the selenide layer, and the extra Li ions distribute in the interstitial sites. In either way, the gating process is irreversible when the $V_g$ is swept backwards from 5.29 V to -2 V, so that the new structure cannot transform back to the (Li,Fe)OHFeSe phase evidenced by the fact that the *d*-value of the (001) reflection for the new structure shifts from 11.02 Å to 9.91 Å. This is the reason why the device shows an insulating behavior without ferromagnetism as shown in figs. S5C and S5D of the supplementary materials.

In order to understand the new crystal structure and reversibility during the Li gating process in (Li,Fe)OHFeSe thin flake, we perform the first-principle calculations to search for the most stable atomic configurations and crystalline structures. We find that Li ions driven into thin flake by electric field prefer to replace Fe in (Li,Fe)OH layer and expel Fe out of the (Li,Fe)OH layer to the interstitial sites. The Fe ions expelled from the hydroxide layers can migrate to the center of the Se square of FeSe layer, and the optimized stable structure with $c = 10.18$ Å is obtained for the new structural phase with $Fe_{0.2}LiOHFeSe$ as shown in Fig. 3D. According to our NEB (nudged elastic band) searching calculation, the potential barrier for the Fe ions expelled by Li ions from the hydroxide layer to the interstitial site of FeSe layer is about 0.16 eV, and it is about 0.45 eV for the reversal process as shown in the fig. S7 of the supplementary materials. The process with such a low potential barrier is easily reversible. On the other hand, we also explore the possibility of Li substitution for the Fe in the selenide layer after Fe ions in (Li,Fe)OH layer are completely replaced by the Li ions. We consider two situations: 1) Li occupation in interstitial site, and 2) Li substitution for Fe in the FeSe layers. We find that the total energy of the latter is about 2.3 eV higher than the former, ruling out the possibility of Li substitution for Fe in the FeSe layers at high gating voltage. In addition, more Li ions will occupy the interstitial sites between the LiOH and $Fe_{1+x}Se$ layers with further increasing Li ions, and block the migration path of the Fe at the center positions of the Se square sublattice back to LiOH layer. Therefore, once the applied gate voltage is switched off, the Fe could not return back to the original positions in LiOH layer completely, so that the gating process becomes irreversible with a large expansion of the $c$-axis lattice parameter as observed in Fig. 3B. In one word, the theoretical calculations not only qualitatively explain the experimental observations, but also predict the structure of the novel metastable crystalline phase.

In summary, we can control the crystal structural transformation with electric field through the recently developed SIC-FET, and the corresponding electronic and magnetic states are also regulated simultaneously. High-$T_c$ superconductivity and ferromagnetism rarely exist simultaneously in the same phase diagram of a solid, and it is thus striking to realize the control of phase transition by electric field from high-$T_c$ superconducting state to ferromagnetically insulating state. Our study demonstrates the potential applications of the SIC-FET for multifunctional devices and its superior tunability for an electronic system that transcends the ability of carrier doping. Our findings open a new way for the control of magnetic and electronic states, and pave a way to access the metastable phases and to find the unexpected physical properties.

**Acknowledgements**

This work is supported by the National Key R&D Program of the MOST of China (Grant Nos. 2017YFA0303001), the National Natural Science Foundation of China (Grants Nos. 11534010), and the Strategic Priority Research Program (B) of the Chinese Academy of Sciences and the Key Research Program of Frontier Sciences CAS (Grant No. QYZDY-SSW-SLH021)

**Author contributions**

X.H.C. conceived and coordinated the project, and is responsible for the infrastructure and project direction. L.K.M., B.L., N.Z.W. contributed equally to this work. L.K.M., L.B., and N.Z.W. performed device fabrication and measurements with assistance from C.S., F.B.M., Z.L.S., J.H.C., C.S.Z., Z.S.. K.S.Y, D.Y.L. and L.J.Z performed theoretical analysis and calculations. L.K.M., B.L., N. Z. W., T.W. and X.H.C. analyzed the data. X.H.C., B.L, L.K.M, N.Z.W and Z.S. wrote the paper. All authors discussed the results and commented on the manuscript.

**Additional information**

The authors declare no competing financial interests. Correspondence and requests for materials should be addressed to X.H.C. (chenxh@ustc.edu.cn).

**Figure caption**

**Figure 1. Resistance of (Li,Fe)OHFeSe controlled by gate voltage with SIC-FET device.**

**(A):** A schematic view of the (Li,Fe)OHFeSe based SIC-FET device. From the bottom to the top: Cr/Au back gate layer, finely polished Li ion conductor substrate, (Li,Fe)OHFeSe thin flake and Cr/Au *Hall* bar electrodes. **(B):** Gate voltage dependent resistance of a (Li,Fe)OHFeSe thin flake with thickness of ~ 120 nm in SIC-FET device. The continuously swept gate voltage is applied at $T = 260$ K with a scan rate of 1 mVs$^{-1}$. The inset shows the optical image of a (Li,Fe)OHFeSe thin flake with a standard *Hall* bar configuration, and the current and voltage terminals are labeled. **(C):** $T_c$ gradually increases with increasing gate voltage. The inset is the magnified view of the superconducting transition at the optimal $T_c$. **(D):** With further increasing gate voltage, $T_c$ decreases gradually, and eventually the sample becomes an insulator.

**Figure 2. The evolution of $R_{xy}$(B) with different gate voltage at $T = 60$ K, and $R_{xy}$(B) and $MR$(B) as a function of temperature at $V_g = 5.15$ V. (A)-(C):** The magnetic field dependent *Hall* resistance $R_{xy}$ at 60 K. $R_{xy}$(B) shows linear behavior in the superconducting regime. When the sample is tuned into insulating regime, an obvious anomalous *Hall* effect is observed. **(D) and (E):** The magnetic field dependent *Hall* resistance $R_{xy}$ and magnetoresistance ($MR$) at different temperatures. The $MR$(B) curves are shifted vertically with a step of 5% for clarity.

**Figure 3.** *In-situ* **X-ray diffraction patterns of (Li,Fe)OHFeSe thin flake at various gate voltages in the SIC-FET device. (A):** The XRD patterns of the (Li,Fe)OHFeSe thin flake with thickness of 175 nm under different $V_g$ at $T$ = 150 K. The XRD pattern taken at $T$ = 300 K before gating is also plotted in the figure. **(B):** The magnified view of the angle range from 7° to 11°. The corresponding $c$-axis lattice parameters at some gate voltages are labelled beside the diffraction peak. **(C):** The gate voltage dependent resistance of (Li,Fe)OHFeSe thin flake in the SIC-FET device at $T$ = 260 K with the highest $V_g$ in the irreversible regime. **(D):** Calculated crystalline lattice configurations with $a = b$ = 8.1854 Å, $c$ = 10.1792 Å (space group: $P4/n$) for the $Fe_{0.2}$LiOHFeSe phase.

**Figure 4. The phase diagram of the gate-tuned (Li,Fe)OHFeSe thin flake.**

The $T_c$ at different gate voltages is determined by the onset critical temperature $T_c^{onset}$ from the resistance measurements. The $c$-axis lattice parameter is determined by the position of (00l) peaks from the in-situ XRD measurement. Once the superconductivity is completely suppressed, the system goes into the regime of ferromagnetic insulator. The dashed line at $V_g$ = 5.18 V indicates the boundary between the reversible and irreversible gating regimes.

**Figure 1.**

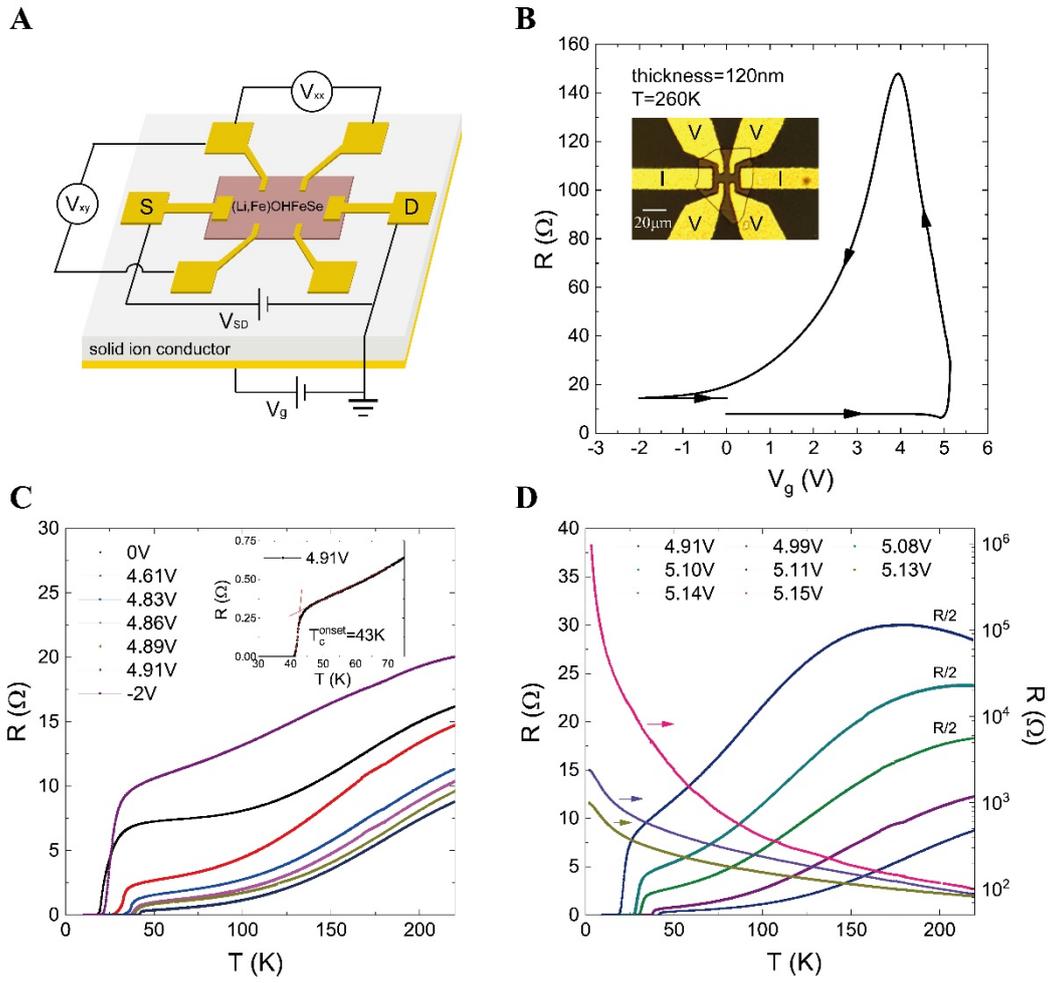

**Figure 2.**

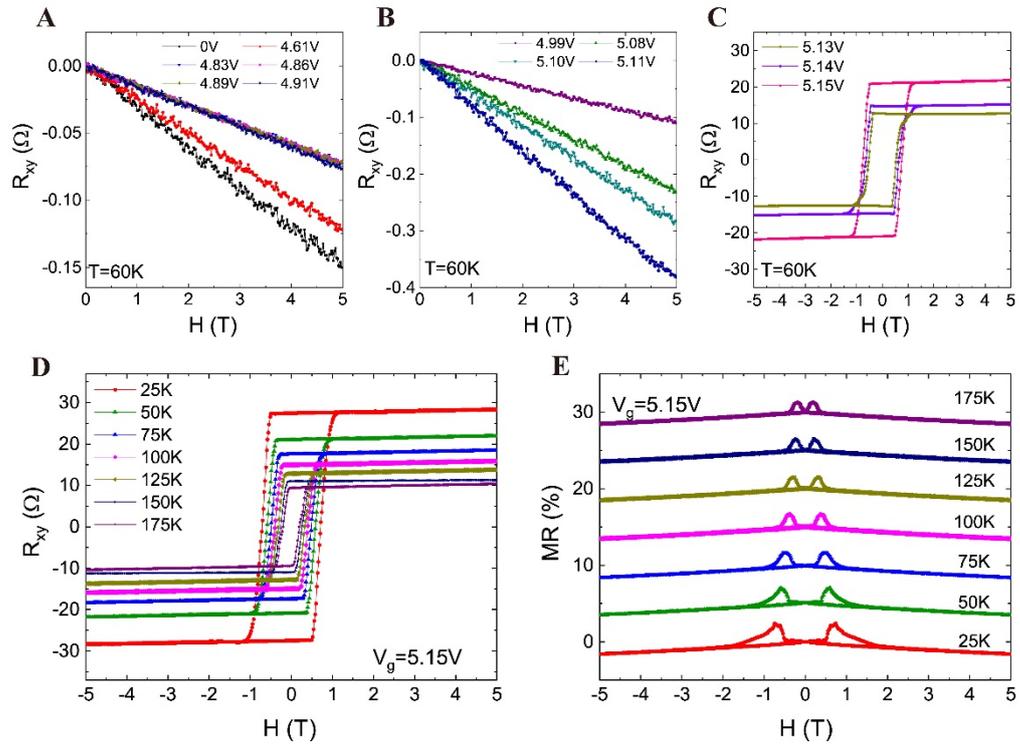

**Figure 3.**

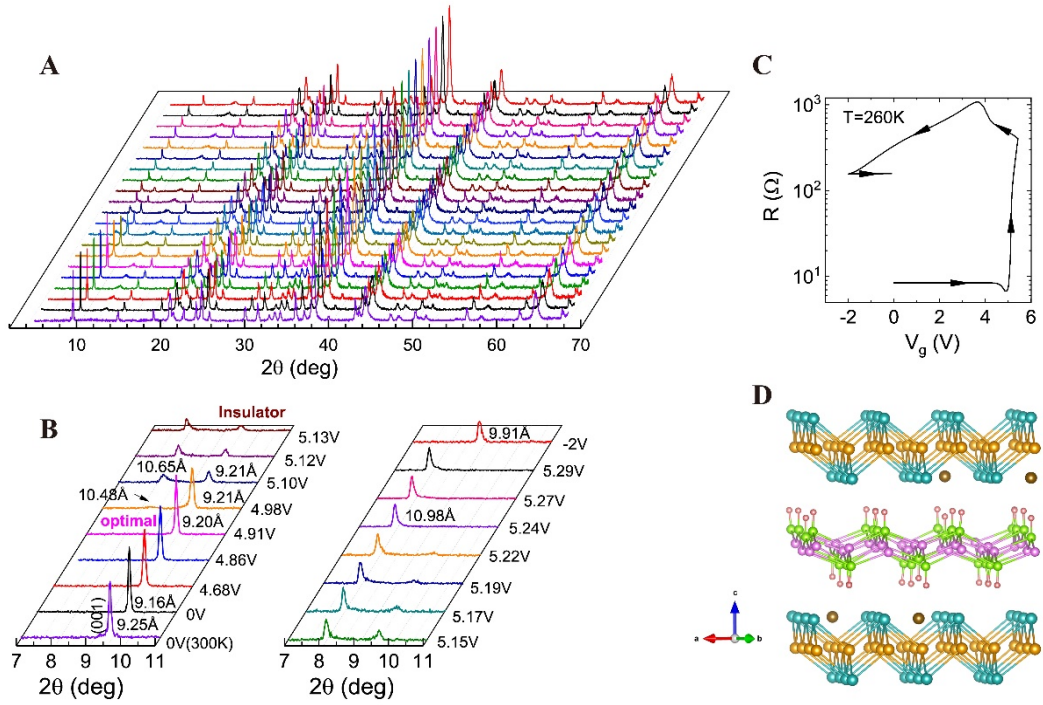

**Figure 4.**

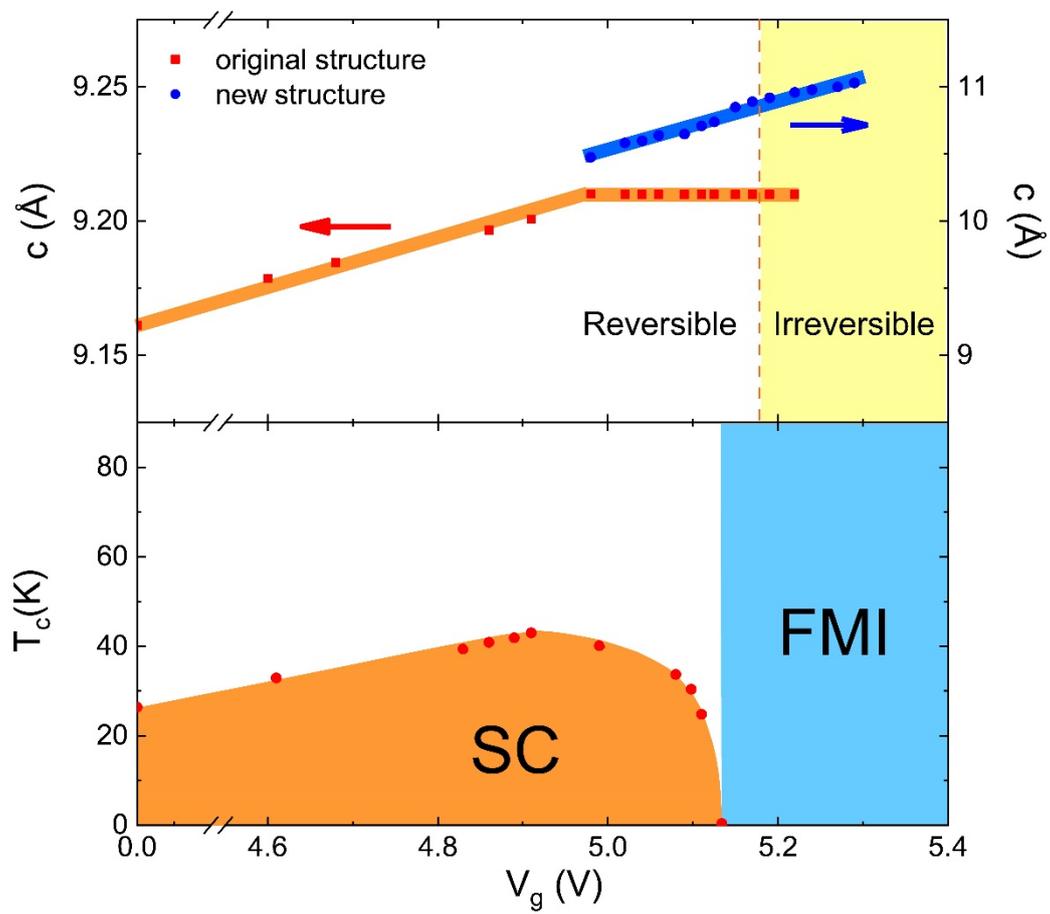